\documentclass{INTERSPEECH2023}

\interspeechcameraready

\usepackage{amsmath, amssymb, booktabs, multirow, graphicx,cite, algpseudocode, algorithm, tabularx, color}

\usepackage{subcaption}
\usepackage{hyperref}
\usepackage{pgfplots}
\pgfplotsset{compat=1.18}
\usepackage{bm}

\title{Random Utterance Concatenation Based Data Augmentation for \\ Improving Short-video Speech Recognition}
\name{Yist Y. Lin$^{*}$, Tao Han$^{*}$, Haihua Xu$^{*}$, Van Tung Pham, Yerbolat Khassanov,\\ Tze Yuang Chong,  Yi He, Lu Lu, Zejun Ma}
\address{ByteDance Research}
\email{\{yist.lin0, tao.han, haihua.xu, van.pham, yerb.khass, tychong, heyi.hy, lulu0314, mazejun\}@bytedance.com}
\begin{document}
%
\maketitle
\def\thefootnote{*}\footnotetext{Authors equally contributed  to this work.}\def\thefootnote{\arabic{footnote}}

\begin{abstract}
\vspace{-1mm}
One of limitations in end-to-end automatic speech recognition (ASR) framework is its performance would be compromised if train-test utterance lengths are mismatched. 
In this paper, we propose an 
on-the-fly random utterance concatenation (RUC) based data augmentation method to alleviate train-test utterance length mismatch issue for short-video ASR task. 
Specifically, we are motivated by observations that our human-transcribed training utterances tend to be much shorter for short-video spontaneous speech ($\sim$3 seconds on average), while our test utterance generated from voice activity detection front-end is much longer ($\sim$10 seconds on average). 
Such a mismatch can lead to suboptimal performance.
Empirically, it's observed the proposed RUC method significantly improves long utterance recognition without performance drop on short one.
Overall, it achieves 5.72\% word error rate reduction on average for 15 languages and improved robustness to various utterance length.


\end{abstract}
\vspace{-2mm}

\section{Introduction} \label{sec:intro}
\vspace{-1mm}
End-to-end (E2E) automatic speech recognition (ASR)~\cite{graves2012sequence,chan2016listen,vaswani2017attention,gulati2020conformer} framework has been predominant in both academic and industrial areas\cite{he2019streaming,sainath2020streaming,li2022language}, thanks to its simplicity, compactness, as well as  efficacy in modeling capacity.
However, there still remains unresolved problems for E2E ASR framework. 
One of the problems is the learned E2E ASR models tend to overfit to the utterances that have been seen during the training. This makes the models cannot well generalize to the unseen utterances during the inference.
Here, the so-called ``unseen" can refer to various noise conditions~\cite{kim2017joint, sun2019adversarial}, out-of-vocabulary or tail words~\cite{zheng2021oov, cal2020tail}, as well as sequence length mismatch problems~\cite{arun2019-google,2019c3-google,zhou2019improving,2020c3-google,zlu2021google-input}, to mention a few.

To alleviate the overfitting problem, many effective solutions have been proposed.
One of the earlier pioneer works is dropout~\cite{srivastava2014dropout,dahl2013improving,moon2015rnndrop,cheng2017exploration}, which randomly drops some of the visible or hidden nodes during training. 
To date, dropout is widely employed in deep learning to yield robust networks that can alleviate generalization problems. 
Following that, another popular technique developed to address train-test data mismatch problem is scheduled sampling~\cite{bengio2015scheduled,zhou2019improving,chiu2018state}, which is crucial for autoregressive attention-based encoder-decoder (AED) framework~\cite{chorowski2015attention,chan2016listen,bahdanau2016end}. 
This is because ground-truth tokens are used as prior knowledge during training stage, while during evaluation the prior tokens might be erroneous. 

Another group of works are focused on front-end data augmentation~\cite{ko2015audio,ko2017study,INTERSPEECH-SpecAug-2019,rosenberg2019speech,nguyen2020onetheflyda}. To let models see diverse data, speed perturbation~\cite{ko2015audio}, as well as artificial reverberant noise corruptions~\cite{ko2017study} are applied. Inspired by ``cutout" in computer vision~\cite{devries2017cutout}, the recently proposed SpecAugment~\cite{INTERSPEECH-SpecAug-2019}, which masks a small portion of Mel coefficients along both spectral and temporal axes during training, has demonstrated significant efficacy for ASR performance.

More recently, the problem of train-test utterance length mismatch has drawn increased attention in both machine translation (MT)~\cite{provilkov2021multi} and 
ASR communities~\cite{arun2019-google,2019c3-google,2020c3-google,zlu2021google-input}. 
Specifically, it is observed that the models trained with short utterances can result in significantly degraded results (a lot of deletions) when the input test utterances are much longer. 
To address this problem in MT, \cite{provilkov2021multi} proposed a multi-sentence resampling (MSR) method to concatenate sentences, yielding improved translation performance.
In ASR task, a series of recipes, such as passing random state to simulate utterance concatenation~\cite{arun2019-google}, attention mechanism manipulation~\cite{chorowski2015attention,2019c3-google}, and overlapping decoding~\cite{arun2019-google,2020c3-google}, have been proposed to improve long utterance recognition.

In this paper, we propose a random utterance concatenation (RUC) method to address the train-test utterance length mismatch problem in short-video speech recognition scenario\footnote{Short video refers to the one with duration of around three minutes in this paper.}.
We are motivated from the observation that our human transcribed utterances are too short ($\sim$3 seconds on average), while the test utterances from the front-end voice activity detection (VAD) are much longer ($\sim$10 seconds on average).
The RUC method augments the training data by randomly concatenating two or more utterances.
The intention here is to improve the robustness of ASR system by letting the E2E ASR model see more diverse utterances in length.

Our main contributions lie in the following aspects: First to the best of our knowledge, we are the first to explicitly apply  on-the-fly RUC for ASR training, and reveal the effectiveness of the  proposed method as a front-end data augmentation under ASR scenario.
Secondly, we validate the efficacy of the proposed method on the challenging short-video ASR tasks of 15 languages, where the datasets are real, spontaneous, and noisy, and the amount of the training data is in the range from $\sim$1,000 to $\sim$32,000 hours.
Besides, we have conducted extensive experiments for optimal concatenation settings, as well as a series of analysis under a wide range of train-test length mismatch conditions.
\vspace{-2mm}

\section{Relation to prior work}\label{sec:prior}
\vspace{-1mm}
To alleviate the train-test utterance length mismatch problem in MT task, \cite{provilkov2021multi} proposed a MSR-based sentence concatenation
method that is very close to the proposed RUC here.
Although some analyses are  conducted on Librispeech data for the ASR task, no actual experiment indicating the efficacy for the ASR task has been performed in \cite{provilkov2021multi}. 
 Besides, compared with the MT task, utterance concatenation for the ASR task is  more complicated, since the concatenated utterances are not only potentially heterogeneous in semantics, they might also be acoustically irrelevant. One cannot know for sure whether it is working without empirical study.

Train-test utterance length mismatch problem has been extensively studied in ASR community~\cite{arun2019-google,2019c3-google,2020c3-google} recently. However, the hallmark of the work is to deal with very long-form speech to decode. 
For instance, the length of their target utterances can be as long as several minutes.
Here, our target test utterances are shorter than 25 seconds.
For such utterances,
 the effectiveness of their methods has not been reported. 
Besides, all experiments in \cite{arun2019-google,2019c3-google,2020c3-google} are conducted on RNN-T ASR framework, while our ASR models belong to the AED.
Additionally, the proposed RUC is purely a data augmentation method, and it is orthogonal to the prior approaches proposed in~\cite{arun2019-google,2019c3-google,2020c3-google} that mainly make efforts on model level.

\vspace{-2mm}

\section{Random utterance concatenation}\label{sec:msr}
\vspace{-1mm}
The purpose of the proposed RUC method is to generate longer utterances on-the-fly by randomly 
concatenating  utterances during training. 
Algorithm~\ref{alg:ruc} reveals the implementation details.  
\begin{algorithm}[htbp]
\caption{On-the-fly Random Utterance Concatenation}\label{alg:ruc}
\begin{algorithmic}[1]
\State{$S\leftarrow$ overall training steps}
\State{$B \leftarrow$ batch size of training data}
\State{$N\leftarrow$ maximum number of utterances to concatenate}
\For{$s \leftarrow 1 $ to $S$}
\State{$D\leftarrow$ randomly subset buffer of overall training set}
    \State{$b \leftarrow \varnothing$ }
    \While{$| b | < B$}
        \State{$n\leftarrow$ random integer from 1 to $N$}
        \State{new\_trans $\leftarrow \varnothing$}
        \State{new\_feat $\leftarrow \varnothing$}
        \For{$i \leftarrow 1$ to $n$}
            \State{(transcript, feature) $\leftarrow \text{random\_sample}(D)$}
            \State{new\_trans $\leftarrow$ new\_trans.concat(transcript)}
            \State{new\_feat $\leftarrow$ new\_feat.concat(feature)}
        \EndFor
        \State{b.append((new\_trans, new\_feat))}
    \EndWhile
    \State{do\_one\_step\_training(b)}
\EndFor
\end{algorithmic}
\end{algorithm}
In Algorithm~\ref{alg:ruc}, 
the length of concatenated utterances are constrained within 300 tokens, and the maximum duration is  25 seconds in time. 
The ``feature" in Algorithm~\ref{alg:ruc} can be either raw waveform or MEL spectral coefficients. 
In this paper, all concatenations are performed on MEL feature level. 

The whole RUC training is divided into two stages. The first stage is a normal cross-entropy training with learning rate decay method for 200k steps. After that, we employ the RUC method to fine-tune the model with constant learning rate for another 50k steps.
\vspace{-2mm}

\section{Dataset}\label{sec:data-set}
\vspace{-1mm}
We employ the anonymized short-video dataset to verify the efficacy of the RUC approach for ASR training.
The data is rather challenging. 
Not only are they spontaneous, the genres are also highly diverse.
For instance, within a normal short video, majority of audios are not speech, but music, ambient noise, and other non-verbal sounds, such as laughter, coughs, breath, etc.
Taking German and Swedish as examples, we observe over 90 percent of videos contain only around 3-5 short utterances that are transcribable. 

Table~\ref{tab:train-data-dist} reports the  overall training data statistics for 15 languages used in this work.
The dataset lengths are in the range from 1k to 32k hours. The smallest training dataset is Swedish, with 1k hours. Following Swedish, Japanese and Korean are also not that large, with 1.7k and 2.5k hours respectively.  While the largest  is   English that covers the data from different countries, including US, UK,  Canada, Australia, New Zealand, and South African. 
\begin{table}[htb]
    \centering
    \caption{Training data statistics of 15 languages}~\label{tab:train-data-dist}
    \begin{tabular}{lcc}
    \toprule
        Language (ID) & Hours (K) &Utterances (M) \\
        \midrule
        Burmese (my) & 3.8 & 2.8\\
        Dutch (nl) &4.8& 5.2 \\
        Filipino (fil) & 5.0& 4.8\\
        French (fr) & 7.0& 8.4\\
        German (de) & 9.5& 10.6\\
        Indonesian (id) &9.8 & 9.4\\
        Italian (it) & 9.6& 7.9\\
        Japanese (ja) &1.7& 2.2 \\
        Korean (ko) &2.5& 2.2\\
        Polish (pl) & 9.6&9.5 \\
        Portuguese (pt) &10.0& 13.2 \\
        Russian (ru) &9.0& 6.2\\
        Swedish (sv) &1.0& 1.1\\
        Vietnamese (vi) & 9.7 &  11.7\\
        English (en) & 32.7 & 24.4 \\
        \bottomrule
    \end{tabular}
\end{table}
Table~\ref{tab:train-test-stat} further presents the average utterance length statistics of training and test sets in terms of both time and token units for all languages.
\begin{table}[htb]
    \centering
    \footnotesize
    \renewcommand\arraystretch{1}
    \setlength{\tabcolsep}{0.75mm}
    \caption{Average utterance length statistics (mean $\pm$ standard deviation) in training and test sets for 15 languages}~\label{tab:train-test-stat}
    \begin{tabular}{p{0.55cm}rp{0.175cm}lrp{0.175cm}p{0.75cm}rp{0.175cm}lrp{0.175cm}l}
    \toprule
    \multirow{2}{*}{ID} &\multicolumn{6}{c}{Training} &\multicolumn{6}{c}{Test} \\
	                     & \multicolumn{3}{c}{Duration (s)} & \multicolumn{3}{c}{\#Tokens} & \multicolumn{3}{c}{Duration (s)} & \multicolumn{3}{c}{\#Tokens} \\
	                     \midrule
     my  & 4.62 & $\pm$ & 3.49   & 13.47 & $\pm$ & 11.87 & 10.67 & $\pm$ & 4.51  & 29.2 & $\pm$ & 18.9\\
     nl  & 3.19 & $\pm$ & 2.63   & 9.12 & $\pm$ & 7.76   & 10.93 & $\pm$ & 4.72  & 24.2 & $\pm$ & 17.1\\
     fil & 3.51 & $\pm$ & 2.82   & 9.88 & $\pm$ & 8.46   & 9.98  & $\pm$ & 4.26  & 21.3 & $\pm$ & 13.3\\
     fr  & 2.82 & $\pm$ & 2.44   & 11.30 & $\pm$ & 10.08 & 10.47 & $\pm$ & 4.64  & 34.1 & $\pm$ & 23.1\\
     de  & 3.08 & $\pm$ & 2.86   & 9.74 & $\pm$ & 9.44   & 10.22 & $\pm$ & 4.21  & 26.7 & $\pm$ & 15.9\\
     id  & 3.57 & $\pm$ & 3.15   & 9.09 & $\pm$ & 8.23   & 11.35 & $\pm$ & 4.54  & 20.8 & $\pm$ & 13.9\\
     it  & 4.14 & $\pm$ & 3.22   & 11.53 & $\pm$ & 10.11 & 10.85 & $\pm$ & 4.71  & 26.6 & $\pm$ & 17.5\\
     ja  & 2.63 & $\pm$ & 2.56   & 14.95 & $\pm$ & 14.58 & 12.39 & $\pm$ & 4.11  & 78.0 & $\pm$ & 39.6\\
     ko  & 3.89 & $\pm$ & 3.28   & 8.95 & $\pm$ & 8.59   & 10.33 & $\pm$ & 4.80  & 17.7 & $\pm$ & 13.9\\
     pl  & 3.50 & $\pm$ & 2.88   & 11.81 & $\pm$ & 10.39 & 10.41 & $\pm$ & 4.61  & 32.7 & $\pm$ & 20.9\\
     pt  & 2.73 & $\pm$ & 2.08   & 8.76 & $\pm$ & 7.09   & 10.07 & $\pm$ & 3.97  & 29.8 & $\pm$ & 15.7\\
     ru  & 5.11 & $\pm$ & 3.99   & 14.45 & $\pm$ & 12.52 & 11.11 & $\pm$ & 4.54  & 23.9 & $\pm$ & 15.4\\
     sv  & 3.35 & $\pm$ & 2.70   & 10.30 & $\pm$ & 9.06  & 10.59 & $\pm$ & 4.58  & 29.0 & $\pm$ & 19.4\\
     vi  & 2.88 & $\pm$ & 2.58   & 12.47 & $\pm$ & 11.28 & 10.70 & $\pm$ & 4.73  & 35.8 & $\pm$ & 25.5\\
     en  & 4.84 & $\pm$ & 4.23   & 15.64 & $\pm$ & 15.29 & 11.74 & $\pm$ & 3.73  & 39.8 & $\pm$ & 17.9\\
    \bottomrule
    \end{tabular}
\end{table}
From Table~\ref{tab:train-test-stat}, we can see that the average utterance length of our training data is much shorter than the test data in terms of both time and token units. 
Taking time unit for instance, Russian training set has the longest utterances, 5.11s  on average, and with the ratio of test-train utterance duration  $r\approx2.17$.
Japanese has the shortest utterances,  with 2.63s  on average  and $r\approx 4.71$.
The length mismatch here mainly attributes to two reasons.
On the one hand, as was mentioned earlier, our short-video data is mostly non-speech, and human transcribers tend to have short utterances to transcribe for training data preparation. 
On the other hand, it is related with our specific test settings.
Our incoming test data is an entire short video, and we rely on a NN-based VAD front-end to  first remove non-speech parts, then split longer speech segments into smaller ones. Finally, they are fed to the ASR engine.
To avoid intra-speech segmentation or speech missing,
the VAD tends to output longer speech to the ASR engine, which results in train-test length mismatch.

\vspace{-2mm}
\section{Experiments and results}\label{sec:exp}
\vspace{-1mm}
\subsection{Modeling}
\label{sub:model}
The E2E ASR models employed in this work are based on attention-based encoder-decoder architecture, which are similar to the ones used in~\cite{liu2022ilme}.
Specifically, the encoder is Transformer, while the decoder is LSTM. Transformer's \{layer, dim, head\} parameters are \{18, 512, 8\}, and the feed-forward network dimension is set to 2048 with the GLU activations. The LSTM decoder has four layers with 1024 cells per layer. For robust training, we employed both variational noise (VN)~\cite{graves2013speech} and SpecAugment~\cite{park2019specaugment} methods. The VN training is activated after 10k steps. The SpecAugment is activated after 2k steps, and its frequency \{F, m$_F$\} and temporal \{W, m$_T$, p\} parameters are set to \{27, 2\} and \{100, 1, 0.1\} respectively. 
During  inference, the beam size is fixed to 10, and the best length normalization factor~\cite{wu2016google} is selected for each language independently in [0.0, 0.8] range.
All ASR models employ word piece model~\cite{sennrich-etal-2016-neural,kudo2018sentencepiece} with the vocabulary size being 3-7k.
\vspace{-2mm}

\subsection{Results}
\label{sub:results}
\vspace{-1mm}
Table~\ref{tab:results} reports the WER  and corresponding WER reduction (WERR) results (for N=\{1, 4, 6, 8\}) using the proposed RUC method on 15 languages. In table~\ref{tab:results}, we report 4 categories of experiments, where N=\{4, 6, 8\} means we randomly concatenate up to 4, 6, 8 utterances for the proposed method. As a contrast, N=1 refers to a continual training without any utterance concatenation. As shown in Table~\ref{tab:results}, continual training without the proposed method gets only 0.62\% WERR on average, while the proposed RUC data augmentation method makes consistent significant WERR improvement with three concatenation settings. The general trend is shown to be the larger N the bigger WERR. Specifically the WERR  is from 4.24\% up to 5.72\% on average.
\begin{table}[htb]
    \centering
    \caption{WER (\%)  and corresponding WERR (\%) for 15 languages using the proposed RUC data augmentation method, where N=1 means 50k steps of continual training, and no utterance concatenation being performed}
    \label{tab:results}
    \begin{tabular}{lcrrrrrr}
    \toprule
    ID & Baseline (WER) & N=1 & N=4 & N=6 & N=8 \\
    \midrule 
    my  & 20.65 & -4.00 & 5.94  & 8.36  & 10.15 \\
    nl  & 23.90 & -2.30 & 2.73  & 1.94  & 1.10 \\
    fil & 26.27 &  1.54 & 2.74  & 2.17  & 2.98 \\
    fr  & 19.35 &  1.29 & 5.13  & 6.08  & 6.87 \\
    de  & 15.05 &  4.25 & 7.73  & 9.59  & 10.12 \\
    id  & 21.79 &  2.94 & 5.03  & 6.06  & 5.58 \\
    it  & 18.27 &  2.38 & 4.18  & 3.85  & 3.56 \\
    ja  & 17.66 & -2.89 & -1.45 & 3.57  & 4.10 \\
    ko  & 18.70 &  2.73 & 1.52  & 4.10  & 5.42 \\
    pl  & 13.12 &  3.16 & 0.08  & -1.42 & 1.30 \\
    pt  & 12.72 &  1.77 & 9.07  & 8.94  & 10.06 \\
    ru  & 19.22 & -3.59 & -0.40 & 1.37  & 0.68 \\
    sv  & 24.88 &  2.97 & 5.08  & 1.80  & 1.65 \\
    vi  & 26.73 &  1.48 & 17.23 & 22.38 & 22.53 \\
    en  & 9.62  & -2.39 & -1.04 & -0.49 & -0.31 \\
    \midrule
    Avg.& 19.20 & 0.62  &  4.24 & 5.22  & 5.72 \\
    \bottomrule
    \end{tabular}
\end{table}
In Table~\ref{tab:results}, the RUC has not achieved similar WERRs for each language.
The best WERRs are obtained on Vietnamese (vi), Portuguese (pt), German (de), and Burmese (my) 4 languages, and their best WERRs are over 10\%.
Interestingly, it gains 22.53\% WERR on Vietnamese.
We hypothesize such a huge WERR could be related with too short utterances ($\sim 2.88s$) in the training dataset.
On the opposite side, there are almost no WERRs for English (en), Russian (ru), Polish (pl), and Dutch (nl) 4 languages, and the average utterance length of these 4 languages are over 3.0s.
These might partially mean RUC is potential useful for the training dataset that has a lot of shorter utterances.

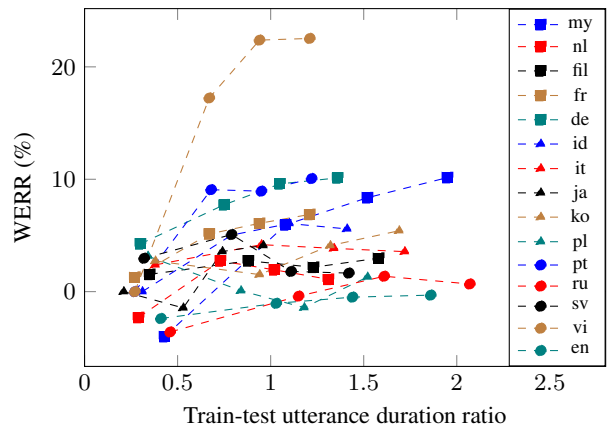
\begin{figure}[htb]
    \centering
    \begin{tikzpicture}
        \begin{axis}[
            height=180, width=240,
            xmin=0, xmax=2.8,
            xlabel={Train-test utterance duration ratio},
            ylabel={WERR (\%)},
            ylabel near ticks,
            legend style={
                anchor=south east,
                at={(axis description cs:1,0)},
                nodes={scale=0.8, transform shape}
                }
            ]
            \addplot[mark=square*, blue, dashed] table { 
                x y
                0.43 -4.0
                1.08 5.94
                1.52 8.36
                1.95 10.15
                };
            \addplot[mark=square*, red, dashed] table { 
                x y
                0.29 -2.30
                0.73 2.73
                1.02 1.94
                1.31 1.10
                };
            \addplot[mark=square*, black, dashed] table { 
                x y
                0.35 1.54
                0.88 2.74
                1.23 2.17
                1.58 2.98
                };
            \addplot[mark=square*, brown, dashed] table { 
                x y
                0.27 1.29
                0.67 5.13
                0.94 6.08
                1.21 6.87
                };
            \addplot[mark=square*, teal, dashed] table { 
                x y
                0.30 4.25
                0.75 7.73
                1.05 9.59
                1.36 10.12
                };
            \addplot[mark=triangle*, blue, dashed] table { 
                x y
                0.31 0
                0.79 5.03
                1.10 6.06
                1.41 5.58
                };
            \addplot[mark=triangle*, red, dashed] table { 
                x y
                0.38 2.38
                0.95 4.18
                1.34 3.85
                1.72 3.56
                };
            \addplot[mark=triangle*, black, dashed] table { 
                x y
                0.21 0
                0.53 -1.45
                0.74 3.57
                0.96 4.10
                };
            \addplot[mark=triangle*, brown, dashed] table { 
                x y
                0.38 2.73
                0.94 1.52
                1.32 4.10
                1.69 5.42
                };
            \addplot[mark=triangle*, teal, dashed] table { 
                x y
                0.34 3.16
                0.84 0.08
                1.18 -1.42
                1.52 1.30
                };
            \addplot[mark=*, blue, dashed] table { 
                x y
                0.27 0
                0.68 9.07
                0.95 8.94
                1.22 10.06
                };
            \addplot[mark=*, red, dashed] table { 
                x y
                0.46 -3.59
                1.15 -0.40
                1.61 1.37
                2.07 0.68
                };
            \addplot[mark=*, black, dashed] table { 
                x y
                0.32 2.97
                0.79 5.08
                1.11 1.80
                1.42 1.65
                };
            \addplot[mark=*, brown, dashed] table { 
                x y
                0.27 0
                0.67 17.23
                0.94 22.38
                1.21 22.53
                };
            \addplot[mark=*, teal, dashed] table { 
                x y
                0.41 -2.39
                1.03 -1.04
                1.44 -0.49
                1.86 -0.31
                };
            \legend{
                my,
                nl,
                fil,
                fr,
                de,
                id,
                it,
                ja,
                ko,
                pl,
                pt,
                ru,
                sv,
                vi,
                en
            }
	\end{axis}
    \end{tikzpicture}
    \caption{WERR vs. train-test utterance duration ratio ($R$) for all languages}
    \label{fig:werr_vs_train_length}
\end{figure}

Figure ~\ref{fig:werr_vs_train_length} plots the WERR versus train-test utterance duration ratio for each of 15 languages.
Denoting $R$ as train-test average utterance duration ratio, we can observe from Figure~\ref{fig:werr_vs_train_length}  that when $R\in [0.5, 1.0]$, all language ASR models obtain increased WERRs except for Korean, and Polish ASR models. Particularly, the latter shows clear performance drop trend.
Interestingly, as $R \in [1.0, 2.0]$, there is no obvious WERR deterioration for any language, and on the contrary, the best performing languages in Table~\ref{tab:results} still get further WERR.
The result in Figure~\ref{fig:werr_vs_train_length} indicates employing RUC to produce training utterances as one to two times long as test utterances ($R$ lies in [1.0, 2.0]) could yield improved ASR performance.

\section{Analysis}\label{sec:analysis}
Train-test mismatch is unavoidable in real applications.
To simulate such a mismatch, we have attempted different test utterance length by VAD, namely, 15s, 12s, 10s, 7s\footnote{$N$s denotes the average utterance length produced with a VAD setting is $N$ seconds.}, so as to examine the robustness of the proposed method.
We find after RUC training the performance fluctuation is considerably reduced.
Table~\ref{tab:wer_diff_vad} reveals the regarding efficacy via a contrast between ASR models with RUC and without RUC training on 4 languages.
\begin{table}[htb]
    \centering
    \setlength{\tabcolsep}{0.75mm}
    \footnotesize
    \caption{WER (\%) under different utterance length by VAD settings and their standard deviations (SD) before and after the proposed RUC training for 4 languages}
    \label{tab:wer_diff_vad}
    \begin{tabular}{lcccccccccc}
    \toprule
    \multirow{2}{*}{ID} & \multicolumn{5}{c}{Without RUC} & \multicolumn{5}{c}{With RUC} \\
    \cmidrule(lr{1mm}){2-6}\cmidrule(lr{1mm}){7-11}
    & 15s & 12s & 10s & 7s & SD & 15s & 12s & 10s & 7s & SD \\
    \midrule
    pt  & 16.61 & 16.54 & 15.72 & 15.37 & \bf 0.61 & 16.02 & 16.12 & 15.72 & 15.59 & \bf 0.25 \\
    vi  & 19.76 & 20.24 & 17.21 & 15.12 & \bf 2.38 & 15.47 & 15.61 & 13.62 & 13.68 & \bf 1.09 \\
    ja  & 15.04 & 15.11 & 13.93 & 13.40 & \bf 0.84 & 13.59 & 13.57 & 12.96 & 13.66 & \bf 0.33 \\
    ko  & 12.19 & 12.27 & 11.93 & 12.31 & \bf 0.17 & 11.93 & 11.77 & 11.82 & 12.24 & \bf 0.21 \\
    \bottomrule
    \end{tabular}
\end{table}
We can observe from Table~\ref{tab:wer_diff_vad} not only is WER performance improved but the WER standard deviations (SDs) with different VAD settings are also remarkably reduced respectively.
We note that the WER SDs of the Korean (ko) are changed little since its WERs between different VAD settings are already rather small, and it gets minor increased after RUC training, from 0.17 to  0.21.

\begin{figure}[htb]
    \centering
    \begin{subfigure}[b]{0.49\linewidth}
    \centering
        \begin{tikzpicture}
            \begin{axis} [
                width=130, ymin=8, ymax=32,
                symbolic x coords={en,pl,ru,nl},
                xtick=data,
                every x tick label/.append style={font=\footnotesize},
                ylabel={WER (\%)}, ylabel near ticks]
                \addplot [
                    only marks,
                    mark=-,
                    mark options={blue, mark size=2pt, line width=2pt}]
                plot [
                    error bars/.cd,
                    y dir=both,
                    y explicit,
                    error mark options={rotate=90, red, mark size=4pt, line width=1pt}]
                table [y error plus=wer-max, y error minus=wer-min] {
                    language y wer-max wer-min
                    en 9.98 0.46 0.36
                    pl 13.67 0.79 0.55
                    ru 19.57 0.52 0.35
                    nl 26.66 5.03 2.76
                };
            \end{axis}
        \end{tikzpicture}
        \caption{Without RUC}
        \label{fig:length_norm_least_wo_ruc}
    \end{subfigure}
    \hfill
    \begin{subfigure}[b]{0.49\linewidth}
        \centering
        \begin{tikzpicture}
            \begin{axis} [
                width=130, ymin=8, ymax=32,
                symbolic x coords={en,pl,ru,nl},
                xtick=data, every x tick label/.append style={font=\footnotesize}]
                \addplot [
                    only marks,
                    mark=-,
                    mark options={blue, mark size=2pt, line width=2pt}]
                plot [
                    error bars/.cd,
                    y dir=both,
                    y explicit,
                    error mark options={rotate=90, red, mark size=4pt, line width=1pt}]
                table [y error plus=wer-max, y error minus=wer-min] {
                    language y wer-max wer-min
                    en 9.74 0.11 0.09
                    pl 13.27 0.21 0.16
                    ru 19.56 0.42 0.43
                    nl 24.73 2.18 1.19
                };
            \end{axis}
        \end{tikzpicture}
        \caption{With RUC}
        \label{fig:length_norm_least_w_ruc}
    \end{subfigure}
    \par\medskip
    \begin{subfigure}[b]{0.49\linewidth}
        \centering
        \begin{tikzpicture}
            \begin{axis} [
                width=130, ymin=11, ymax=46,
                symbolic x coords={pt,de,my,vi},
                xtick=data, every x tick label/.append style={font=\footnotesize},
                ylabel={WER (\%)}, ylabel near ticks]
                \addplot [
                    only marks,
                    mark=-,
                    mark options={blue, mark size=2pt, line width=2pt}]
                plot [
                    error bars/.cd,
                    y dir=both,
                    y explicit,
                    error mark options={rotate=90, red, mark size=4pt, line width=1pt}]
                table [y error plus=wer-max, y error minus=wer-min] {
                    language y wer-max wer-min
                    pt 13.95 2.58 1.23
                    de 17.30 2.93 2.25
                    my 26.47 6.87 5.80
                    vi 35.53 10.16 8.80
                };
            \end{axis}
        \end{tikzpicture}
        \caption{Without RUC}
        \label{fig:length_norm_best_wo_ruc}
    \end{subfigure}
    \hfill
    \begin{subfigure}[b]{0.49\linewidth}
        \centering
        \begin{tikzpicture}
            \begin{axis} [
                width=130, ymin=11, ymax=46,
                symbolic x coords={pt,de,my,vi},
                xtick=data, every x tick label/.append style={font=\footnotesize}]
                \addplot [
                    only marks,
                    mark=-,
                    mark options={blue, mark size=2pt, line width=2pt}]
                plot [
                    error bars/.cd,
                    y dir=both,
                    y explicit,
                    error mark options={rotate=90, red, mark size=4pt, line width=1pt}]
                table [y error plus=wer-max, y error minus=wer-min] {
                    language y wer-max wer-min
                    pt 12.00 0.79 0.48
                    de 13.93 0.62 0.46
                    my 22.17 4.73 3.70
                    vi 24.20 4.22 3.80
                };
                \addlegendentry{\scriptsize{Mean}}
                \addlegendimage{color=black, mark=|, mark options={red, mark repeat=2, mark phase=1, mark size=4pt, line width=1pt}}
                \addlegendentry{\scriptsize{Max \& min}}
            \end{axis}
        \end{tikzpicture}
        \caption{With RUC}
        \label{fig:length_norm_best_w_ruc}
    \end{subfigure}
    \caption{Illustration of RUC efficacy on length normalization for inference through contrasts between those with and without RUC training.}
    \label{fig:length_norm}
\end{figure}
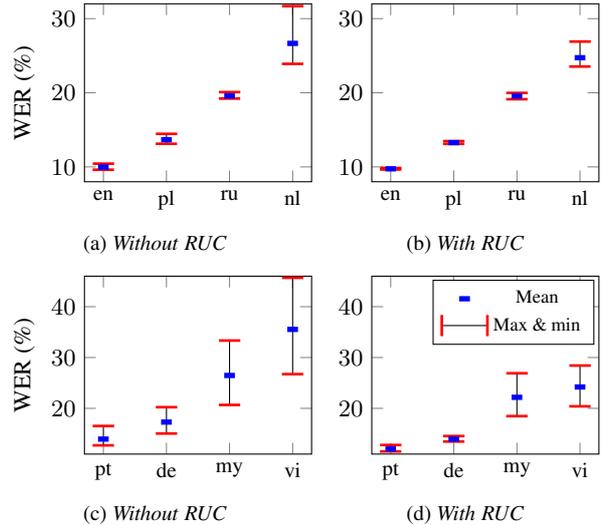

One of the bad effects of train-test utterance length mismatch, particularly training utterance being much shorter than test ones, is that many more deletion errors will arise during inference~\cite{provilkov2021multi}.
To alleviate deletion errors, one would simply employ length normalization
therapy which basically awards longer utterances. We use the length normalization method as advocated in \cite{wu2016google} in this work, and it is formulated as follows:
\begin{align}
    & s(\bm{y}, \bm{x}) = \sum_{i=1}^{m} \log P(y_i\ |\ y_{1:i}, \bm{x}) \label{eq:asr_score} \\
    & s'(\bm{y}, \bm{x}) = s(\bm{y}, \bm{x}) \mathbin{/} \frac{(5 + |\bm{y}|)^\alpha}{(5 + 1)^\alpha} \label{eq:length_normalized_asr_score}
\end{align}
where  $|\bm{y}|$ refers to the token length of the decoded hypothesis, and $\alpha$ is the so-called length normalization  hyper-parameter that controls the dynamic range of the denominator, normally in the range of [0.0, 0.8] in our work.
Eq.~\ref{eq:asr_score} is the normal AED decoding formula, while Eq.~\ref{eq:length_normalized_asr_score} is the implementation formula of Eq.~\ref{eq:asr_score} in practice.
We found that after RUC training, the ASR model is less reliant on normalization, that is,  we found the WER gaps between different normalization factors are much smaller after RUC training.
Figure~\ref{fig:length_norm} illustrates the WER dynamic range on two groups of languages. One group of languages that obtains the least WERR from RUC training are English, Polish, Russian, and Dutch (Figure~\ref{fig:length_norm_least_wo_ruc} and \ref{fig:length_norm_least_w_ruc}), while the other group that has achieved best WERR are Portuguese, German, Burmese, and Vietnamese (Figure~\ref{fig:length_norm_best_wo_ruc} and \ref{fig:length_norm_best_w_ruc}) respectively.
From Figure~\ref{fig:length_norm}, we notice that the WER change dynamics are obviously smaller after RUC training compared with the case of which no RUC training is performed on either group of languages.
This indicates RUC training brings more robust ASR models, which would be beneficial when our ASR engine receives diverse incoming speech.

\begin{figure}[htb]
    \centering
    \begin{tikzpicture}
        \begin{axis}[
            width=160, height=130,
            xlabel={Utterance length (token)},
            ylabel={WER (\%)},
            ytick={10,15,20,25,30,35},
            grid=major,
            ylabel near ticks]
            \addplot[color=red, mark=*] coordinates {
                (10, 34)
                (20, 17.5)
                (30, 13.5)
                (40, 12.3)
                (50, 12.5)
                (60, 11.8)
                (70, 13.3)
                (80, 14.8)
                (90, 18)
                };
            \addplot[color=blue, mark=triangle*] coordinates {
                (10, 33.7)
                (20, 16.7)
                (30, 12.7)
                (40, 11)
                (50, 10.5)
                (60, 10)
                (70, 10.1)
                (80, 9.9)
                (90, 9.8)
                };
            \legend{Without RUC, With RUC}
	\end{axis}
    \end{tikzpicture}
    \caption{Illustration of WER versus test utterance length with and without the proposed RUC method on Swedish.}~\label{fig:ana-wer-vs-test}
    \vspace{-6mm}
\end{figure}
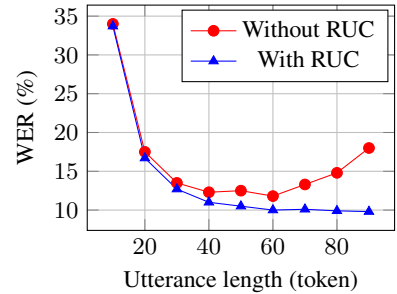

Table~\ref{tab:results} has clearly revealed the effectiveness of the proposed RUC training on overall WERR, and now we are curious about if such a performance improvement  occurs on the overall test utterances with diverse utterance length distribution. To take a closer look at what has happened with more details, we perform RUC on a predefined Swedish \texttt{dev} set, and compare the results of different ASR models with or without RUC training. Figure~\ref{fig:ana-wer-vs-test} plots the details of WER versus utterance length on Swedish.
From Figure~\ref{fig:ana-wer-vs-test}, it is interesting to note that the RUC training method can yield comparable WER results with the normal training method on shorter test utterances whose length is below 20 tokens.
As test utterances are getting longer and longer, the gap between the proposed and conventional methods appears and tends to enlarge.
More specifically, the WER gaps begin to appear for utterances in 20-40 token area, and the proposed RUC method prevails.
As the utterance length is in 40-60 token area, the gap enlarges obviously.
Finally, when test utterance length is over 60 tokens, the normal ASR models without RUC method yields deteriorating WER results, while the RUC method consistently gets improved WERR.
This further demonstrates the effectiveness of the proposed RUC data augmentation method since it is beneficial to the overall test set with wide distribution of utterance length.

\vspace{-2mm}
\section{Conclusion}\label{sec:con}
\vspace{-1mm}
In this work, we proposed an on-the-fly random utterance concatenation method as front-end data augmentation for improving short-video speech recognition.
Specifically, the proposed method addresses train-test utterance length mismatch originated from the situation in which incoming test utterance length is much longer than that in the training set.
We demonstrated its efficacy using diverse ASR models from 15 languages whose datasets are in the range of 1k to 32k hours. 
With the method, we achieved up to 5.72\% WER reduction on average for the overall languages.
By further analysis, we found the proposed data augmentation method can make the ASR model less sensitive to length normalization, which potentially proves that the ASR models are more robust to diverse environments.
Moreover, the proposed method is beneficial to long utterance decoding without any performance drop on short utterance.

\vfill\pagebreak
\clearpage



\bibliographystyle{IEEEtran}
\bibliography{refs}

\end{document}